\documentstyle[aps,prl,preprint,floats,epsfig]{revtex}

\newcommand{\Dst}{$D^{*}$}
\newcommand{\Dstpl}{$D^{*+}$}

\newcommand{\Dnew}{$D_{1}^{0}$(j=1/2)}
\newcommand{\US}{$\Upsilon$(4S)}
\newcommand{\pim}{$\pi^{-}$}

\begin{document}

\preprint{\tighten\vbox{\hbox{\hfil CLEO CONF 99-6}
}}

\title{Observation of a Broad L=1 $c\overline{q}$ State in 
$B^{-} \rightarrow $\Dstpl \pim \pim at CLEO}  

\author{CLEO Collaboration}
\date{\today}

\maketitle
\tighten

\begin{abstract} 
Using 4.7 $fb^{-1}$ of data taken at CESR at energies at and near the \US~
we have studied the decay
$B^{-} \rightarrow D^{*+}\pi^{-}\pi^{-}$ (and its conjugate).  
We observe a new, broad 
charmed meson state, which we interpret as \Dnew, in its decay 
to $D^{*+}\pi^{-}$.
Our preliminary results indicate the mass and width of this L=1 state to be 
$m = (2461^{+41}_{-34}\pm10\pm32) MeV$ and 
$\Gamma = (290^{+101}_{-79}\pm26\pm36) MeV$, with the third uncertainty
associated with the 
parameterization of the relative strong phases.  In addition
we have measured several new branching fractions of charged $B$ mesons.
All quoted results are preliminary.

\end{abstract}
\newpage

{
\renewcommand{\thefootnote}{\fnsymbol{footnote}}

\begin{center}
S.~Anderson,$^{1}$ V.~V.~Frolov,$^{1}$ Y.~Kubota,$^{1}$
S.~J.~Lee,$^{1}$ R.~Mahapatra,$^{1}$ J.~J.~O'Neill,$^{1}$
R.~Poling,$^{1}$ T.~Riehle,$^{1}$ A.~Smith,$^{1}$
S.~Ahmed,$^{2}$ M.~S.~Alam,$^{2}$
% ...
\end{center}
 
\small
\begin{center}
$^{1}${University of Minnesota, Minneapolis, Minnesota 55455}\\
$^{2}${State University of New York at Albany, Albany, New York 12222}\\
% ...
\end{center}

\setcounter{footnote}{0}
}
\newpage

\section{Charmed Meson Spectroscopy}
	The lowest mass charmed mesons are the pseudoscalar 
$D$ and the vector \Dst~, with '*' denoting
that the spins
of the $c$ and $\overline{q}$ are aligned.  For one unit of
orbital angular momentum between these partons, there are four states of 
positive parity, which carry the labels $D_{1}$ and 
$D_{0}^{*}, D_{1}^{*}, D_{2}^{*}$; these four states are collectively
referred to as $D_{J}$.  

If $m_c << \Lambda_{QCD}$ ({\it i.e.,}
if the charm quark mass is small on the scale of QCD energies), then the
spectroscopy of these L=1 mesons would be a singlet and a triplet, much
as with positronium in QED.  On the other hand, if  $m_c >> \Lambda_{QCD}$
then Heavy Quark Effective Theory (HQET) would predict that the light
quark degrees of freedom decouple from those of the $c$ quark so that the
quantum number governing the spectroscopy would be 
$\vec{j} = \vec{L} + \vec{S_{\overline{q}}}.$  This would result in two
doublets, one with $j = 1/2$ and the other 
with $j=3/2$. As with the hydrogen atom in QED, 
adding in the spin of the heavy partner results in
'hyperfine' splitting within these doublets.
The $c$ quark is not sufficiently massive for HQET to be considered
perfect, but massive enough that the effects of HQET should be evident.
Therefore some mixing is expected between the two states of $J^{P} = 1^{+}$. 

\begin{figure}[htb]
\centerline{\epsfig{file=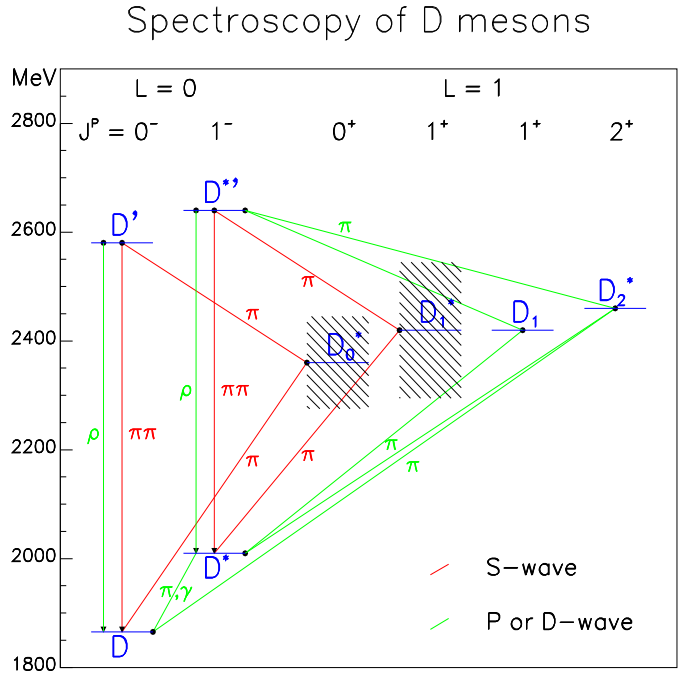,height=3.5in,width=3.5in}}
\vspace{10pt}
\caption{Spectrum of the charmed mesons.  Prior to this analysis the
two shaded L=1 states were unobserved.  HQET would predict these two
states to have $j=1/2$ and the two observed L=1 states to have $j=3/2$.}
\label{spectrum_2}
\end{figure}

Three of the $D_{J}$ can decay to $D^{*}\pi$, as shown in the level diagram
of Figure \ref{spectrum_2}.  
The $D_{2}^{*}$ decay can only be d-wave, whereas conserving
$J$ and $P$ allow the decays of the $1^{+}$ states to be either
s-wave or d-wave.  However, if HQET is invoked, then the $1^{+}$ state
with $j=3/2$ would decay only via d-wave (and therefore be narrow) but
the $j=1/2$ state would be restricted to s-wave decay (and be correspondingly
broader).  Two narrow states {\it have} already been observed \cite{CAE},
the $D_{1}$(2420) and the $D_{2}^{*}$(2460).

\section{Search Procedure and Analysis}

We have used the CLEO detector \cite{CLEONIM} at the Cornell Electron Storage
Ring to search for the $D_{J}$ in the decay chain 
(and its conjugate)
$B^{-} \rightarrow D_{J}^{0} \pi_{1}^{-}$, with 
$D_{J}^{0} \rightarrow D^{*+} \pi_{2}^{-}$ and
$D^{*+} \rightarrow D^{0}\pi_{3}^{+}$ .
The pion subscripts identify
which pion corresponds to which piece in the decay chain.  The data
used correspond to 3.1$fb^{-1}$ taken at the \US~ energy and 1.6$fb^{-1}$
taken at a slightly lower energy 
to model the underlying continuum in our fits.

Using only four-momentum conservation and the measurement
of the momenta of the three pions
gives 28 constraints on the seven-particle system, providing 
a ``0C'' fit.  This ``partial reconstruction'' technique,
in which we do not reconstruct the decay of the $D^{0}$,
greatly enhances our statistics, although it also increases
the overall level of background.

Because both the $B$ and $\pi$
are pseudoscalars, the $D_{J}$ will be produced in a totally
aligned state,
so that the angular information in the decays will help
sort among the various resonant contributions.
Using $B$ decays gives us four variables to use in our maximum likelihood
fitting procedure.  The first of these is the angle between the momenta
of $\pi_{2}$ and $\pi_{1}$ in the $D_{J}$ rest frame; these two vectors
also define a plane denoted $\epsilon_{12}$.  The second variable,
similarly, is the angle between the momenta 
of $\pi_{3}$ and $\pi_{2}$ in the $D^{*+}$ rest frame; these vectors
form the plane $\epsilon_{23}$.  The angle between the planes
$\epsilon_{12}$ and $\epsilon_{23}$ is the third variable and the
$D^{*+}\pi_{2}$ invariant mass, which will be the mass of the
$D_{J}$ candidates, is the fourth.

Backgrounds fall into three categories in this analysis.  The off-\US~
running gives us the shape 
and expected yield
of the {\it continuum background}, i.e.,
events that do not have $B\overline{B}$ parentage.  Standard
CLEO Monte Carlo simulations, with equivalent luminosity of several
times the data, provide the shape of {\it generic B background}.  Finally,
there are several $B$ decay channels that have kinematics similar to
that of the signal, forming {\it correlated backgrounds} that have
high efficiency.  
These include
decay chains that do involve a $D_{J}$ but with an incorrect pion in the
reconstruction and are handled by specialized, high-statistics
simulations.

The data are then fit\cite{Rodriguez} 
via maximum likelihood using the four
variables described above, to these expected background shapes plus an 
overall amplitude
function that has three Breit-Wigner distributions 
for the three $D_{J}$ that
decay via a $D^{*}$ and a non-resonant 
$B^{-}\rightarrow D^{*+}\pi^{-}\pi^{-}$ contribution.  The two $1^{+}$
states are allowed to mix, each having angular contributions 
for both s-wave and d-wave decay. In addition we try several
parameterizations that allow for 
strong interaction phase differences among the decay amplitudes.

\section{Results}

\begin{figure}[htb]
\centerline{\epsfig{file=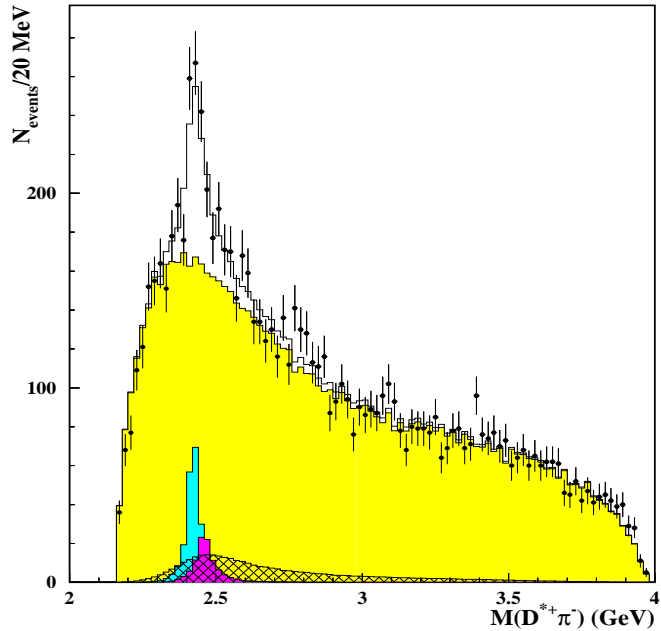,height=3.5in,width=3.5in}}
\vspace{10pt}
\caption{The distributions of data (points with error bars), total
expected background (lightly shaded histogram), 
and best overall fit (open histogram)
in the $D^{*+}\pi^{-}$ invariant mass.  Also shown are the three
resonant structures from the fit minimization - the narrow, previously
established $1^{+}$ and $2^{+}$ states and the new, broad state.}
\label{mDstpi}
\end{figure}

The $m(D^{*}\pi)$ distribution of the data is shown in Fig.\ref{mDstpi},
along with the fit projections 
and the sum of all the expected backgrounds.  In
addition to the
two established narrow resonances, there is clearly a need for the third,
broad resonance which is taken to be the $D_{1}^{0}$(j=1/2) state.
The fit also indicates there is a small non-resonant contribution.

The broad state, which we identify as the $D_{1}^{0}$(j=1/2),
has mass and width of:

\begin{equation}
m = (2461~^{+41}_{-34}~\pm10~\pm32)~{\rm MeV};~~~~~ 
\Gamma = (290~^{+101}_{-79}~\pm26~\pm36)~{\rm  MeV}.
\end{equation}

%
%
The three listed uncertainties are associated with statistics, general 
systematics ({\it e.g.}, selection criteria, mass and width of 
the known, narrow resonances), and the parameterization of the strong
phases in the decay.  Using efficiencies determined from simulation,
the fit naturally gives us the product branching fractions
$\mathcal{B}$($B^{-} \rightarrow D_{J}^{0} \pi^{-}$)$\cdot$
$\mathcal{B}$($D_{J}^{0} \rightarrow D^{*+} \pi^{-}$), which are shown in 
Table~\ref{table:1} as $\mathcal{B}$$_{B}\cdot$$\mathcal{B}$$_{D_J}$.  
Based on Clebsch-Gordon coefficients and other
theoretical inputs\cite{theory} about the $D_{J}$ decays, we also
show the branching fractions $\mathcal{B}$($B^{-} \rightarrow D_{J}^{0} \pi^{-}$),
which, at 1 to 1.5 per mil, are somewhat larger than expected\cite{BrFr}.

\begin{table}[htb]
\caption{Preliminary results form the 4D maximum likelihood fit. Shown are the
product branching fractions and $B$ decay branching fractions, as 
described in the text, and the significance of that decay chain in the
overall fit to the data.}
\label{table:1}
\begin{tabular}{lcclll}
\hline
Resonance 		& Status  &  Width   & 
$\mathcal{B}$$_{B}\cdot$$\mathcal{B}$$_{D_J} (10^{-4}$)&
	$\mathcal{B}$$_{B} (10^{-3})$	&  Significance \\
\hline
$D_{1}$(2420)	& Known	& Narrow  & 
   $6.9^{+1.8}_{-1.4}\pm1.1\pm0.4$  & $1.04 \pm 0.33$ & $>4.5\sigma$ \\
$D_{1}$(j=1/2)	& New	& Broad   & 
   $10.6 \pm1.9\pm1.7\pm2.3$        & $1.59 \pm 0.52$ & $>5.5\sigma$ \\
$D_{2}$(2460)	& Known	& Narrow  & 
   $3.1\pm0.8\pm0.4\pm0.3$	    & $1.55 \pm 0.49$ & $>4.5\sigma$ \\
non-resonant	&	&	  &
		                    & $0.97 \pm 0.44$ & $>2.0\sigma$\\
\hline
\end{tabular}
\end{table}

\end{document}